\documentclass[lettersize,journal]{IEEEtran}
\usepackage{amsmath,amsfonts}
\usepackage{algorithmic}
\usepackage{algorithm}
\usepackage{array}
\usepackage[caption=false,font=normalsize,labelfont=sf,textfont=sf]{subfig}
\usepackage{textcomp}
\usepackage{stfloats}
\usepackage{url}
\usepackage{verbatim}
\usepackage{graphicx}
\usepackage{cite}
\usepackage{booktabs}
\usepackage{multirow}
\usepackage{mdframed}
\usepackage{tipa}
\begin{document}

\title{Seq2seq for Automatic Paraphasia Detection in Aphasic Speech}
\author{Matthew Perez, Duc Le, Amrit Romana, Elise Jones, Keli Licata, Emily Mower Provost}


\maketitle

\begin{abstract}
Paraphasias are speech errors that are often characteristic of aphasia and they represent an important signal in assessing disease severity and subtype.
Traditionally, clinicians manually identify paraphasias by transcribing and analyzing speech-language samples, which can be a time-consuming and burdensome process.
Identifying paraphasias automatically can greatly help clinicians with the transcription process and ultimately facilitate more efficient and consistent aphasia assessment.
Previous research has demonstrated the feasibility of automatic paraphasia detection by training an automatic speech recognition (ASR) model to extract transcripts and then training a separate paraphasia detection model on a set of hand-engineered features.
In this paper, we propose a novel, sequence-to-sequence (seq2seq) model that is trained end-to-end (E2E) to perform both ASR and paraphasia detection tasks. 
We show that the proposed model outperforms the previous state-of-the-art approach for both word-level and utterance-level paraphasia detection tasks and provide additional follow-up evaluations to further understand the proposed model behavior.
\end{abstract}


\section{Introduction}
\IEEEPARstart{A}{phasia} is a common language disorder that occurs as a result of damage to the brain and can ultimately impair the communication abilities (both expressive and receptive) of an individual. Aphasia affects over two million people in the United States and nearly 180,000 acquire aphasia each year following a medical event such as a traumatic brain injury or stroke~\cite{aphasiaweb}.
Aphasia can manifest in a variety of ways that can negatively impact speech production. 
One example of this is through increased speech errors, such as paraphasias.

Paraphasias are a type of communication error and identifying paraphasias can aid clinicians in characterizing an individual's aphasia and developing targeted intervention strategies~\cite{spreen2003assessment}. 
In this work, we focus on identifying phonemic and neologistic paraphasias~\cite{helm2004manual,SALING200731}.
\begin{itemize}
    \item \emph{phonemic} paraphasias involve substituting, omitting, or rearranging phonemes (i.e. `shut' $\rightarrow$ `zut')
    \item \emph{neologistic} paraphasias involve substituting a nonsensical word in place of the target word (i.e. `bottle' $\rightarrow$ `flibber')
\end{itemize}
Clinical research has highlighted the impact that accurate paraphasia detection plays in predicting recovery patterns and guiding treatment planning~\cite{fergadiotis2016algorithmic,mckinnon2018types}. 
Importantly, automated tools that effectively identify the presence of paraphasias in an individual's spoken output can ultimately allow for more efficient and consistent assessment procedures for language disorders like aphasia.

Previous automatic paraphasia detection work has used a pipeline consisting of an automatic speech recognition (ASR) model, transcript-derived feature extraction, and then a paraphasia classification model~\cite{le2017paraphasia}.
Although the authors demonstrated the feasibility of automatic paraphasia detection, the pipeline required three separate processes that had to be trained/computed independently.
In this work, we propose learning all these pipeline components within a single model that is trained E2E.

We present a novel framework for automatic paraphasia detection that uses a sequence-to-sequence (seq2seq) model to perform both ASR and paraphasia detection tasks.
We first acknowledge that paraphasia detection systems are largely dependent on ASR performance and evaluate several E2E ASR architectures (including seq2seq). 
We then evaluate the proposed seq2seq model on a paraphasia detection task and compare against the previous state-of-the-art (SOTA) approach.
We investigate the effects of single-task learning (STL) and multi-task learning (MTL) objectives on the proposed seq2seq model and further analyze model performance using additional word-level paraphasia detection metrics such as temporal distance and time tolerant recall to supplement our findings.
Lastly, we analyze the effect of tokenizer size on paraphasia detection and present some example transcriptions from the model to highlight some of the strengths and limitations of the proposed approach.
The research contributions of this paper are:
\begin{itemize}
    \item An evaluation of the seq2seq architecture for automatic aphasic speech recognition.
    \item The proposed seq2seq model for E2E paraphasia detection.
    \item Assessment of the impact of pretraining on ASR and paraphasia detection tasks.
    \item The effects of single-task (STL) and multi-task (MTL) learning objectives on the proposed seq2seq paraphasia detection model.
    \item An analysis of hyperparameters such as tokenizer size on the proposed model performance.
    \item An analysis and discussion of sampled output from the proposed paraphasia detection model.
\end{itemize}

\section{Background}
\subsection{Aphasia Assessment}
Traditional aphasia assessment practices involve tasks designed to elicit spontaneous speech-language samples, such as those involving descriptions of multi-action pictures or responses to conversational questions~\cite{mayer2003functional,Prins,jaecks2012diagnosing}.
These samples are obtained, transcribed, and analyzed by a speech-language pathologist (SLP) and the resulting analyses can be used for aphasia classification~\cite{risser1985western,fraser2014automated}, treatment planning~\cite{prins2004analyzing}, and progress monitoring~\cite{grande2008basic}.
Ultimately, both transcription and analysis can consume a lot of a SLP's already limited time.
Machine learning systems that can automatically analyze aphasic speech can greatly aid SLPs with the aphasia assessment process and allow for more time to be devoted to patient contact and care.
One form of analysis that can be improved with machine learning is automatic paraphasia detection.
Clinical works have shown that paraphasias are a useful biomarker in characterizing different types of aphasia and ultimately assessing aphasia severity~\cite{spreen2003assessment,goodglass1997word,mckinnon2018types}.
Ultimately, identifying certain types of paraphasia can greatly aid SLPs with aphasia assessment efforts and the development of targeted intervention strategies.

\subsection{Aphasia Treatment Planning}
Improving assessment procedures by introducing the use of automated tools can also facilitate treatment planning and support a SLP's selection of the most appropriate therapy objectives for each PWA.
Treatment for PWA can include a variety of speech-language therapy approaches targeting functional communication across language domains, including spoken language expression, spoken language comprehension, reading comprehension, and written expression~\cite{denes1996intensive,bhogal2003intensity,abad2013automatic,ballard2019feasibility,asha-aphasia}.
Traditionally, speech therapy involves meeting regularly with a SLP to help manage speech-language difficulties.
However, given the rise of ubiquitous computing and smart devices, some clinical works have explored the use of digital technology to supplement treatment and rehabilitation plans, particularly when traditional in-person therapy services are limited due to geographic or financial constraints.

Ballard et al. investigated app-based speech therapy for those with apraxia of speech and aphasia~\cite{ballard2019feasibility}. 
The app uses ASR to recognize input speech related to a naming task and provide feedback to the user. 
Results showed that participants exhibited increased word production accuracy over time. 
However, one limitation of this app is that there were no means for providing feedback regarding paraphasic errors, which could reinforce error patterns and/or contribute to a PWA's limited awareness of errors~\cite{kurland2014ipractice}.  
For remote speech-language therapy applications, feedback from automatic paraphasia detection can be useful in guiding user-driven intervention.
As SLP's tailor treatment planning to the needs of PWA's, it's important the benefits of using app-based technology with automatic error detection methods to supplement traditional speech therapy approaches.

\subsection{Paraphasia Detection}
Several works have demonstrated the ability to identify paraphasias from text input~\cite{fergadiotis2016algorithmic,pai2020unsupervised}. However, a limitation of these approaches is that they rely on manual transcripts and are ultimately not fully automatic when considering speech as an input signal.

One work, by Le et al. has investigated a fully automatic pipeline for paraphasia detection~\cite{le2017paraphasia}, which relied on a hidden markov model-based Multitask Learning Bidirectional Long Short-Term Memory (MTL-BLSTM) acoustic model to first produce transcriptions. 
From these transcriptions, features such as pronunciation, word and phone durations, and phoneme posterior distance are extracted and used to train a donwstream paraphasia classifier.
The authors used two evaluation schemes: the first is augmented word error rate (AWER), which is a word-level metric used to evaluate both transcription and paraphasia label. 
The second is the average F1-score between the negative and positive paraphasia classes, which is computed at the utterance-level. 
Figure~\ref{fig:text} shows an example of how a transcript is combined with paraphasia labels for AWER evaluation.
The authors present the first results for automatic paraphasia detection on this set achieving AWERs of 53.5, 54.2, and 47.8 and F1 scores of 0.594, 0.611, and 0.604 for phonemic+neologistic, phonemic, and neologistic paraphasia detection, respectively. 
To the best of our knowledge, this work by Le et al. represents the closest approach to ours for automatic paraphasia detection.
In this paper, we focus on improving automatic paraphasia detection using a novel seq2seq model that learns both ASR and paraphasia detection tasks E2E.

\begin{figure*}
\begin{mdframed}
\[
  \begin{array}{l l}
    \text{CHAT Transcript:} & \text{I have efezi\textschwa@u [: aphasia] [* n:k]} \\
    \text{Post-processed Transcript:} & \text{I have efezia} \\
    \text{Paraphasia Labels:} & \text{0 \hspace{2pt} 0 \hspace{15pt}  1} \\
    \text{AWER Transription:} & \text{I/0 have/0 efezia/1}
  \end{array}
  \]
\end{mdframed}
\caption{Example showcasing how text and paraphasia labels are concatenated for AWER evaluation. Paraphasia labels are binary with 0=non-paraphasia and 1=paraphasia.}
\label{fig:text}
\end{figure*}

\subsection{Aphasic Speech Recognition}
ASR often represents an important first step before automatic aphasic analysis such as paraphasia detection and precious works have shown that poor ASR transcription can negatively impact downstream analyses~\cite{fraser2013automatic,le2018automatic}. 
With this in mind, training ASR models that perform well on aphasic speech is critical for automatic paraphasia detection.
In this section, we review ASR research focused on improving aphasic speech recognition.
Previous works have focused on overcoming challenges such as abnormal speech patterns, high speaker variability, and data scarcity~\cite{christensen2014automatic,le2018automatic,perez2020aphasic,green2021automatic}.
These challenges make it difficult to apply or adapt traditional off-the-shelf systems due to the data mismatch that exists between speech from healthy controls, which is typically used to train off-the-shelf systems, and disordered speech~\cite{gutz2022validity}.
With this in mind, many researchers opt to train in-domain ASR models for aphasic speech recognition. However, training custom models using supervised learning techniques is also difficult due to the aformentioned challenges and the scarcity of labeled data for PWAs~\cite{le2018automatic,fraser2013automatic,torre2021improving,lee2016automatic,mahmoud2023comparative}.

Some earlier aphasic research used traditional ASR model frameworks, which have separate acoustic, language, and pronunciation models. These works focused on improving the acoustic model which consisted of a hidden markov model, deep neural network (HMM-DNN)~\cite{le2016improving,le2018automatic,perez2018automatic,perez2020aphasic}.
Previous work by Le et al. has focused on using speaker-embeddings such as i-vectors, out-of-domain training, and a Multitask Learning Bidirectional Long Short-Term Memory (MTL-BLSTM) architecture to improve aphasic speech recognition~\cite{le2016improving,le2018automatic}. The BLSTM layers of this model capture both forward and backward dependencies in the input sequences, allowing for better context modeling. Additionally, multitask learning of both senone and monophone labels allow for additional model regularization and improved performance.

End-to-end (E2E) ASR systems focus on modeling word sequences directly from acoustic frames without the need for an HMM. Additionally, these approaches learn the traditional components of acoustic, language, and pronunciation models all together in a single architecture. 
Some examples of E2E models are Connectionist Temporal Classification (CTC) models or sequence-to-sequence (seq2seq) models.
Generally, these models are transformer-based and have been pretrained on vast amounts of speech data before they are finetuned E2E for aphasic speech recognition.
For example, Torre et al. explored fine-tuning a pretrained ASR model by adding an extra layer and optimizing with CTC loss for multilingual Aphasic speech recognition. 
The pretrained model they used is Wav2Vec2-XLSR and the authors were able to show that this approach outperformed existing HMM-DNN approaches~\cite{torre2021improving} on the Spanish and English corpora for ApashiaBank. 

Seq2seq is another approach for E2E ASR model training and involves an encoder-decoder architecture.
Another approach is to use a seq2seq ASR model, which ignores the frame-independence assumption made by traditional HMM or CTC learning approaches and is able to optimize word error rate more directly~\cite{battenberg2017exploring}. The seq2seq approach models the speech recognition task as a machine translation task, and, especially with the use of transformers, has shown much success on traditional ASR benchmarks~\cite{dong2018speech,vaswani2017attention}. A small body of work has started to investigate seq2seq models for aphasic speech recognition~\cite{tang2023new,peng2023comparative}.
Peng et al. proposed an E-branchformer that achieves strong ASR performance across a variety of datasets~\cite{peng2023comparative}.
Tang et al. illustrated how seq2seq frameworks can benefit from leveraging pretrained, self-supervised models. In their work, they finetune a seq2seq model with a pretrained WavLM model and perform multitask learning with ASR and aphasia detection~\cite{tang2023new}.

In this work, we evaluate some of the methods explored above and investigate the role of pretraining on aphasic speech recognition systems as a means of improving automatic paraphasia detection. 
We then show how a seq2seq model can be trained to consider both ASR and paraphasia detection tasks.

\section{Dataset}
We use the AphasiaBank dataset, which is a collection of multimedia data for the study of communication in aphasia~\cite{macwhinney2011aphasiabank}. The database is collected from multiple institutions and contains data in several languages, however, we specifically use the English audio data from the Protocol and the Scripts (non-protocol) sets. 
The Protocol dataset is composed of both Aphasic and Control data collected from 26 different institutions. The speech data consists of a free-form discussion with an interviewer along with several discourse tasks including free speech, picture descriptions, story narratives, and procedural discourse. 
The Scripts dataset is composed of Aphasic data from the Fridriksson subset and contains speech data consisting of read scripts on different topics (advocacy, eggs, vast, and weather).
The Scripts dataset is particularly useful for paraphasia detection due to its increased frequency of paraphasias, compared to the Protocol dataset, with phonemic and nelogistic paraphasias representing 12\% and 6.5\% of all words.
Participants were assessed using the Western Aphasia Battery - Revised (WAB-R), which is a standard test used for assessing aphasia~\cite{kertesz2007western}. We group PWAs into severity classes based on the WAB-R Aphasia Quotient (AQ) following a similar approach to that outlined in~\cite{le2016automatic,le2018automatic} based on mild, moderate, severe, and very severe.
Table~\ref{tab:demo} contains the total time and the percentage of paraphasias for each dataset and severity class.

\section{Transcript Pre-processing}
\label{sec:transcript-preprocess}
All utterances were transcribed following the CHAT transcription format and included timestamps for both participant and interviewer speech segments~
\cite{macwhinney2014childes}. 
We isolate participant speech and discard utterances that have labelled unintelligible speech or overlapping speech between participant and clinician.

We preprocess both the Protocol and Scripts transcripts following the approach described in ~\cite{le2017paraphasia}. 
Non-word phonological errors are transcribed in the International Phonetic Alphabet (IPA) format and each IPA pronunciation is heuristically mapped to a sequence of phones.
We add additional heuristics that convert this phonemic sequence into a pseuodo-word target. Figure~\ref{fig:text} shows an example of a non-word phonological error `aphasia' becoming `efezia'. 
Lastly, we normalize the transcripts to lowercase and remove punctuations.

\begin{table}[h!]
\centering
\begin{tabular}{| l | l r r |} 
 \toprule
 Dataset & Severity & Time (hrs) & \parbox{2.5cm}{ \raggedleft Paraphasia Token \\ \raggedleft Representation (\%) }\\
 \midrule
  \multirow{6}{*}{Protocol} & Control & 38.3 & 0.00 \\
  & Mild & 36.0 & 0.01 \\
  & Moderate & 19.3 & 0.02 \\
  & Severe & 3.3 & 0.03 \\
  & Very Severe & 0.5 & 0.08 \\
  \cline{2-4}
  \rule{0pt}{2.5ex}
  & Total & 97.4 & 0.03 \\
  \midrule
\multirow{1}{*}{Scripts}
  & Total & 3.0  & 0.24 \\
 \bottomrule
\end{tabular}
\vspace{5pt}
\caption{Dataset Info in Hours}
\label{tab:demo}
\end{table}

\begin{figure*}
  \centering
    \includegraphics[width=\linewidth,height=0.3\textheight,keepaspectratio]{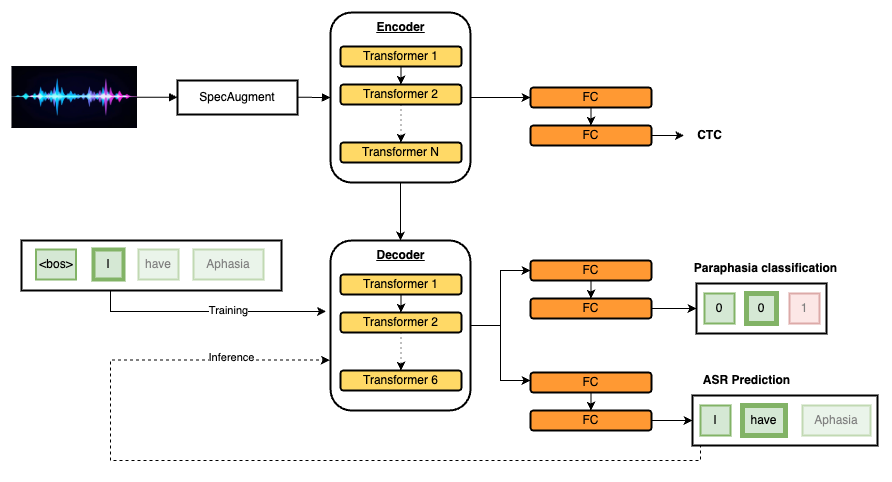}
  \caption{Model Architecture for seq2seq speech recognition. Our best results use a pretrained WavLM model as the encoder.}
  \label{fig:arch}
\end{figure*}

\section{Methods}
\subsection{ASR Models}
ASR is a critical first step in the automatic paraphasia detection pipeline. 
With this in mind, we evaluate a variety of different ASR models, which include HMM-based, encoder-only, and encoder-decoder (i.e. seq2seq) architectures. 
We also, investigate the impact of pretraining on E2E ASR models (i.e. encoder-only and seq2seq) using Wav2Vec2, HuBERT, Whisper, and WavLM models.

\noindent\textbf{Hybrid HMM-BLSTM}\\
Hybrid HMM-DNN acoustic models are considered a more traditional tool for ASR modeling.
Le et al. used a variant of an HMM-DNN, referred to as the MTL-BLSTM, to transcribe aphasic speech before paraphasia detection.
The MTL-BLSTM relied on a manually curated lexicon that is based on the CMU dictionary and contains 39 phones.
The MTL-BLSTM uses mel-filterbank coefficient (MFB) features augmented with utterance-level i-vectors and is optimized to predict both monophone and senone labels. This method decodes the MTL-BLSTM output with a trigram language model that was built on the training set. We implement this approach using pytorch-kaldi~\cite{ravanelli2019pytorch}.


\noindent\textbf{Wav2Vec2.0} \\
Wav2Vec2.0 consists of a CNN-based speech feature encoder, a quantization module, and a transformer network~\cite{baevski2020wav2vec}. The input to Wav2Vec2.0 is raw audio, which is passed through the feature encoder with a receptive field of 25ms and a stride of 20ms. The output of the CNN is then passed to the transformer network which learns contextualized speech representations.
Wav2Vec2.0 was pretrained to learn meaningful representations through a two-step process of extracting pseudo-targets from the audio via a quantization module and then learning to predict these targets with added noise (i.e. masked prediction). 
In our work, we use a large Wav2Vec2.0 model that consists of 317M parameters and was pretrained and finetuned on 960 hours of Libri-light and Librispeech\footnote{https://huggingface.co/facebook/wav2vec2-large-960h-lv60-self}.

\noindent\textbf{HuBERT} \\
HuBERT builds on the initial Wav2Vec2.0 pretraining process. Rather than learning the pseudo-targets while training, these pseudo-targets are created prior to training via k-mean clustering. Additionally, HuBERT uses embeddings from the intermediate layers of the BERT encoder to generate better targets throughout the learning process~\cite{hsu2021hubert}. As a result, HuBERT has been shown to be on par with or better than Wav2Vec2.0 for ASR benchmark tasks~\cite{yang2021superb}.
In our experiments, we use a large HuBERT model that consists of 317M parameters and has been pretrained on LibriLight and finetuned on 960h of Librispeech data\footnote{https://huggingface.co/facebook/hubert-large-ls960-ft}.

\noindent\textbf{WavLM} \\
WavLM extends the HuBERT framework by focusing on data augmentation during pretraining in order to improve speaker representation learning and spoken content modeling~\cite{chen2022wavlm}. This is achieved by introducing denoising as an objective learning task in addition to masked speech prediction. Additionally, WavLM uses a gated relative position bias for the Transformer structure to better capture the sequence ordering of speech input. 
As a result, WavLM generalizes well to many downstream tasks and is currently the top-ranked model on the SUPERB benchmark, which is designed to evaluate universal shared representations for a diverse set of speech processing tasks~\cite{yang2021superb}.
The implementation we use consists of 317M parameters and was trained on 60k hours of LibriLight, 10k hours of GigaSpeech, and 24k hours of VoxPopuli\footnote{https://huggingface.co/microsoft/wavlm-large}.

\noindent\textbf{Whisper} \\
The Whisper architecture consists of an encoder-decoder Transformer that is trained in an E2E fashion~\cite{radford2022robust}. Audio representations are passed to the encoder and the decoder is in charge of predicting text with the inclusion of special tokens designed to direct learning for language identification, phrase-level alignment, and multilingual transcription.
Whisper is trained on a large and diverse dataset and as a result can generalize well to unseen datasets compared to other SSL models.
OpenAI notes that Whisper makes 50\% fewer errors compared to other self-supervised learning (SSL) models for zero-shot learning tasks. We consider Whisper in a zero-shot setting (off-the-shelf application) in our experiments, in addition to fine-tuning this model.
The base model we use is trained on 680k hours of labeled data and has 74M parameters\footnote{https://huggingface.co/openai/whisper-base}. For our off-the-shelf model we use the Whisper-X framework with a whisper base model~\cite{bain2023whisperx}.

\noindent\textbf{Encoder-only} \\
We investigate fine-tuning the pretrained models listed above in an E2E fashion using encoder-only or encoder-decoder (seq2seq) architectures.
For encoder-only finetuning, we add three fully connected layers of sizes 1024, 1024, and 28 to the pretrained model, where 28 is the number of character targets for English. We then train with CTC loss, which learns the alignment between audio frames and characters~\cite{graves2006connectionist}.

\noindent\textbf{Encoder-Decoder} \\
For seq2seq finetuning, we use an encoder-decoder architecture where the encoder is a $N$ layer Transformer, depending on the pretrained model used. If pretraining is not used, then $N$=12.
The decoder is a six-layer transformer, following the default seq2seq architecture used in ~\cite{vaswani2017attention}. 
We use a SentencePiece tokenizer with a unigram tokenization scheme, which is instrumental in breaking down input text into manageable subword units, facilitating more granular and accurate language processing. We use a token size of 500 due to the constrained nature of the AphasiaBank dataset.
The seq2seq model is optimized using a joint CTC-attention loss criterion~\cite{kim2017joint} which is a weighted sum of CTC and CE loss following equation~\ref{seq2seq_loss} where $\alpha$ in our experiments is set to a default value of 0.3.
\begin{equation}
    \mathcal{L}=\alpha*\mathcal{L}_{CTC} + (1-\alpha)\mathcal{L}_{CE}
    \label{seq2seq_loss}
\end{equation}

For both encoder-only and encoder-decoder approaches, we use SpecAugment~\cite{park2019specaugment} to create perturbations in the time domain by resampling utterances at different rates of [0.8, 0.9, 0.95, 1.0, 1.05, 1.1, 1.2]. Work by Green et. al. has suggested that time masking within SpecAugment is more effective for disordered speech recognition than frequency masking\cite{green2021automatic}.

\subsection{Paraphasia Detection Model}
The proposed paraphasia detection model is a transformer-based encoder-decoder architecture and is outlined in Figure~\ref{fig:arch}. 
The encoder output ($H_{enc}$) is fed to a CTC layer.
$H_{enc}$ is also fed to the decoder to predict both subword token $y_t$ and paraphasia label $p_t$ in a sequence.
We employ a loss function based on the negative log-likelihood to compute the cross-entropy loss for ASR prediction and paraphasia classification. Both cross-entropy losses are summed together for a total loss shown in equation~\ref{eq:loss}.
\begin{equation}
\begin{aligned}
    Loss = -logP(y_t|y_1,y_2,...,y_{t-1},x) + \\
    -logP(p_t|y_1,y_2,...,y_{t-1},x)
\end{aligned}
\label{eq:loss}
\end{equation}

During the training phase, subword paraphasia labels are generated by assigning the word-level paraphasia label to each of the word's constituent subwords.
In the inference phase, we aggregate the subword paraphasia labels for each word using an 'OR' function, meaning that if any subword of a word is labeled as a paraphasia, the entire word is classified as a paraphasia.

For our proposed seq2seq model, we investigate both single-task learning (STL) and multitask learning (MTL) objectives. 
The STL objective represents a paraphasia detection pipeline similar to ~\cite{le2017paraphasia} which first optimizes for ASR and then paraphasia detection.
The MTL objective represents learning both of these tasks simultaneously in the network.

For the STL models, we optimize for ASR-only on the Protocol dataset, then finetune using an ASR-only objective on the Scripts dataset for five epochs followed by a paraphasia detection-only objective for the remainder of the training process.
For the MTL models, we first optimize for both ASR and paraphasia detection tasks on the Protocol dataset. 
This involves upsampling the utterances with paraphasias to ensure balanced paraphasia representation in our mini-batches since the protocol dataset has a very limited number of paraphasias compared to the Scripts dataset. 
During this first tuning step on the Protocol dataset we combine both phonemic and neologistic paraphasias into a single class so that the model can learn to detect both types of paraphasias. 
We then finetune using the Scripts dataset and optimize for both ASR and paraphasia detection tasks.
The code for our proposed model can be found at the following github repository~\footnote{\url{https://github.com/matthewkperez/speechbrain_Paraphasia_Detection}}.

\section{Experiments}
\subsection{Aphasia ASR}
\label{experiments:ASR}
We use the Protocol dataset and partition the data into speaker-independent train, dev, and test sets using 70\%, 10\%, and 20\% respectively.
We focus exclusively on the participant speech segments. We remove utterances that are less than 0.75s due to poor alignment and utterances that are greater than 10s due to the hardware constraints of training Transformer-based models. 
All audio data are downsampled to 16kHz data.

We evaluate model performance using word error rate (WER) and provide a detailed breakdown across speaker severity. 
Both the hybrid HMM-DNN and CTC models make use of a trigram language model based on the training set. 
Before decoding, we perform a hyperparameter sweep for controlling smoothing and back-off where $alpha$=[0.4,0.5,0.6] and $beta$=[0.8,1.0,1.2]. 
For seq2seq model decoding, we sweep over the $ctc weight$=[0.2,0.3,0.4]. These hyperparameter sweeps were performed on the validation set and the optimal values were used for decoding on the test set.

\subsection{Paraphasia Detection}
\label{experiments:paraphasia}
For the task of paraphasia detection, we compare our work against that of ~\cite{le2017paraphasia} and investigate the E2E model training for automatic paraphasia detection.
We follow the same training and evaluation scheme as ~\cite{le2017paraphasia} for consistency.
For all models, we follow a two-step training approach that consists of first training on the much larger Protocol dataset, and second finetuning to the domain of Scripts dataset.
We use the same Protocol dataset partitioning outlined in section~\ref{experiments:ASR}.
The Scripts dataset is split into speaker-independent folds and model training and evaluation is performed in a leave-one-subject-out fashion following previous work.
We aggregate our results across all test folds and then compute evaluation metrics following the work of Le et. al. ~\cite{le2017paraphasia}.
Paraphasia detection is treated as a binary classification task, so we train independent models for detecting phonemic (p), neologistic (n), or phonemeic-neologistic (p-n) paraphasias.
We present the paraphasia detection results at both the word- and utterance-levels on the Scripts dataset.

\begin{figure}
    \centering
    \includegraphics[width=\linewidth,height=0.15\textheight,keepaspectratio]{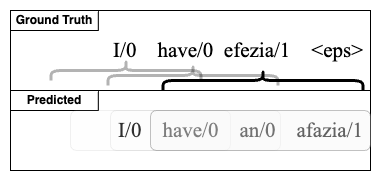}
    \caption{Example of $\omega$=1 is used when computing TTR for a misaligned AWER transcript. The evaluation of the paraphasia label for the word `efezia' with an $\omega$=1 results in a TP whereas an $\omega$=0 results in a FN}
    \label{fig:tp}
\end{figure}

\noindent\textbf{Evaluation Metrics} \\
For word-level evaluation, we use augmented word error rate (AWER) which was previously used by Le et al.~\cite{le2017paraphasia}. 
We create ground truth and predicted AWER transcripts by appending the paraphasia labels to each word in the corresponding transcripts (see Figure \ref{fig:text} for an example), and then calculate AWER as the WER between the ground truth and predicted AWER transcripts. 
AWER represents a high evaluation standard since it requires both the word and paraphasia label to be correctly recognized.
A limitation of this metric is that it does not allow us to focus the evaluation on just word-level paraphasia detection.
To focus our evaluation on just the paraphasia detection performance, we can consider isolating paraphasia labels from the AWER transcripts and comparing the ground truth sequence $Y$=[$y_1$, $y_2$,...,$y_G$], $y\in\{0,1\}$ to the predicted sequence, $\hat{Y}$=[$\hat{y_1}$,$\hat{y_2}$,...,$\hat{y_P}$], $\hat{y}\in\{0,1\}$.
We include two additional word-level metrics that are designed for evaluating fixed-length sequences like word-level paraphasias.


The first metric is temporal distance (TD)~\cite{kovacs2019evaluation}, which is the sum of the target-to-candidate (TTC) and candidate-to-target (CTT) distances, where a lower TD indicates a better score.
The TTC, outlined in equation~\ref{eq:TTC}, is the sum of distances from each target paraphasia ($y_i$) to the closest predicted paraphasia ($\hat{y}_j$).
The CTT, outlined in equation~\ref{eq:CTT}, is the sum of distances from each predicted paraphasia ($\hat{y}_j$) to the closest target paraphasia ($y_i$).
Effectively, TTC punishes false negatives (FN) and CTT punishes false positives (FP). As a result, the effect on TD is that predicted paraphasias that are close in proximity to target paraphasias will result in a good metric.

\begin{equation}
    \text{TD}_{\downarrow}(Y,\hat{Y}) = TTC(Y,\hat{Y}) + CTT(Y,\hat{Y})
\label{eq:TD}
\end{equation}

\begin{equation}
    \text{TTC}(Y,\hat{Y}) = \sum_{i=0}^{G} y_i * \left( \min_{0 \leq j<P}\{|i-j| : y_i == \hat{y}_j\} \right)
\label{eq:TTC}
\end{equation}

\begin{equation}
    \text{CTT}(Y,\hat{Y}) = \sum_{j=0}^{P} \hat{y_j} * \left( \min_{0 \leq i<G}\{|i-j| : y_i == \hat{y}_j\} \right)
\label{eq:CTT}
\end{equation}

The second metric is time tolerant recall (TTR) following equation~\ref{eq:ttr}, which involves computing true positives (TP), false negatives (FN) within a given buffer or window~\cite{scharwachter2020statistical}. Figure~\ref{fig:tp} illustrates an example of this, where the window size is 1.
Equation~\ref{eq:tp} and equation~\ref{eq:fn} show how TP and FN are computed, where $\omega$ is the window size and $I$(c) evaluates to 1 if and only if the condition $c$ is true.
\\

\begin{equation}
    \text{TTR}(Y,\hat{Y}) = \frac{\text{TP}(Y,\hat{Y})} {\text{TP}(Y,\hat{Y}) + \text{FN}(Y,\hat{Y})}
\label{eq:ttr}
\end{equation}
\\

\begin{equation}
    \text{TP}(Y,\hat{Y}) = \sum_{i=0}^{G} y_i * I\left( \sum_{j = \max(0, i-\omega)}^{\min(N, i+\omega)} \hat{y}_j > 0 \right)
\label{eq:tp}
\end{equation}

\begin{equation}
\text{FN}(Y,\hat{Y}) = \sum_{i=0}^{G} y_i * \left( 1- I\left( \sum_{j = \max(0, i-\omega)}^{\min(P, i+\omega)} \hat{y}_j > 0 \right) \right)
\label{eq:fn}
\end{equation}

We also perform utterance-level evaluations using the average F1-score of both the control and paraphasia classes, which was used in prior works~\cite{le2017paraphasia}.

\begin{table}
\centering
\begin{tabular}{ |l || c | c | c | c |  }
 \toprule
 Model  & Mild & Moderate & Severe & Very Severe \\
 \midrule
 MTL-BLSTM~\cite{le2018automatic} & 39.4 & 42.8 & 49.7 & 55.3 \\
 \midrule
 \multicolumn{5}{|c|}{Encoder-only} \\
 \midrule
  Whisper-small* & 32.6 & 40.6 & 43.7 & 65.3 \\
  Whisper-small & 31.1 & 33.8 & 41.5 & 52.1 \\
 Wav2Vec2  & 17.5 & 23.0 & 30.0 & 47.9 \\
 HuBERT  & 16.7 & 22.0 & 28.7 & 50.7 \\
 WavLM  & 16.2 & 22.2 & 28.9 & 46.6 \\
 \midrule
 \multicolumn{5}{|c|}{Seq2Seq} \\
 \midrule
 Transformer-Transformer & 32.9 & 39.5 & 44.7 & 46.6 \\
 Wav2Vec2-Transformer & 17.8 & 27.6 & 32.6 & 69.9 \\
 HuBERT-Transformer  & 16.4 & 24.7 & 29.4 & 61.6 \\
 WavLM-Transformer & \textbf{14.1} & \textbf{20.5} & \textbf{26.6} & \textbf{45.2} \\
 \bottomrule
\end{tabular}
\vspace{5pt}
\caption{WER of ASR models. * indicates off-the-shelf system with no finetuning}
 \label{tab:asr}
\end{table}

\section{Results}
\subsection{ASR}
Table~\ref{tab:asr} shows the WERs for each model we evaluated. We see that an off-the-shelf Whisper model outperforms the previous MTL-BLSTM, achieving WERs of 32.6, 40.6, 43.7, and 65.3 for mild, moderate, severe, and very severe aphasia, respectively. 
This highlights the advantage of using modern architectures that have been pretrained on vast amounts data over traditional HMM-DNN acoustic models.

We find that fine-tuning the Whisper model on the AphasiaBank dataset leads to further improvements over the off-the-shelf model. 
Further, all of the encoder-only models that finetune with CTC loss, outperform the off-the-shelf Whisper model. HuBERT and WavLM achieve the lowest WERs out of these, with HuBERT achieving WERs of 22.0 and 28.7 for moderate and severe aphasia respectively and WavLM achieving WERs of 16.2 and 46.6 for mild and very severe respectively. 
This highlights the benefit of using pretrained speech models for aphasic speech recognition and finetuning them E2E with CTC loss.

Lastly, we find that a seq2seq model with a randomly initialized transformer encoder achieves worse ASR performance than finetuned encoder-only models.
However, when the seq2seq encoder is pretrained, we see performance improvements over the previous approaches, especially when a WavLM model is used.
The WavLM-Transformer model achieves the best performance across all presented methods with WERs of 14.1, 20.5, 26.6, and 45.2 for mild, moderate, severe, and very severe aphasia respectively.
We note that when comparing E2E finetuning approaches (encoder-only and seq2seq) across the pretrained speech models, we see some minor performance differences in WER, specifically with Wav2Vec2 and HuBERT.

These results highlight the importance of leveraging pretrained models for E2E aphasic speech recognition. 
Additionally, finetuning these pretrained models is critical likely due to the issue of data mismatch during pretraining.
Lastly, we saw how model design, pretraining, and optimization choices can impact ASR performance as seen by the differences between CTC models and seq2seq models.
With these design considerations in mind, we will now explore how the seq2seq model can be extended for paraphasia detection.

\begin{table*}
\centering
\begin{tabular}{ |l || c | c | c  || c | c | c  || c | c | c |  }
 \toprule
  \multicolumn{1}{|c||}{} & \multicolumn{6}{|c||}{Word-level} & \multicolumn{3}{|c||}{Utterance-level}\\
  \midrule
 \multicolumn{1}{|c||}{} & \multicolumn{3}{|c||}{AWER} & \multicolumn{3}{|c||}{TD}  & \multicolumn{3}{|c|}{F1-score}\\
 \midrule
 Method & Phn+Neo & Phn & Neo & Phn+Neo & Phn & Neo & Phn+Neo & Phn & Neo \\
 \midrule
Le et. al.~\cite{le2018automatic} & 53.5 & 54.2 & 47.8 &-&-&-& .594 & .611  & .604 \\
 \midrule
STL (proposed) &&&&&&&&& \\
 \hspace{10pt}Wav2Vec2-Transformer &148.7&139.4&120.9& 
 54.3 & 19.9 & 10.3  & 
 .687 & .590 & .636 \\
 \hspace{10pt}HuBERT-Transformer &124.7&157.6&113.2&
 45.2 & 29.2 & 12.1  &
 .691 & .629 & .669 \\
 \hspace{10pt}WavLM-Transformer &117.0&125.4&148.5&
 56.3 & 19.5 & 13.2  &
 \textbf{.706} & .638 & .640 \\
 \midrule
MTL (proposed) &&&&&&&&& \\
 \hspace{10pt}Wav2Vec2-Transformer & 52.0 & 48.9 & 46.7 &
 8.5 & 8.1 & 5.7&
 .693 & .638 & .656 \\
 \hspace{10pt}HuBERT-Transformer & 49.9 & 46.9 & 44.3 &
 \textbf{8.1} & \textbf{8.0 } & 5.5 &
 .703 & \textbf{.643} & .671  \\
\hspace{10pt}WavLM-Transformer & \textbf{48.4} & \textbf{45.0} & \textbf{30.4} &
8.5  & 8.1  & \textbf{5.0} &
.688 & .635 & \textbf{.688}  \\
 \bottomrule
\end{tabular}
\vspace{5pt}
\caption{Paraphasia detection results. Word-level evaluation is measured with AWER. Utterance-level evaluation is measured with F1-score. Results are aggregated over all speaker-independent folds.}
 \label{tab:para}
\end{table*}

\subsection{Paraphasia Detection}
\label{sec:exp-PD}
Starting first with the word-level metrics, we can see that in table~\ref{tab:para}, the seq2seq models trained with STL have very large AWERs. 
We note that this is due to the E2E model setup and that once the learning objective switches from ASR to paraphasia detection, ASR performance begins to decrease since it is no longer explicitly optimized.
However, we see that when MTL is used and both tasks are jointly optimized the resulting models have much lower AWERs compared to STL models. 
The best-performing model in terms of AWER is the MTL WavLM-Transformer, which provides significant performance improvements over the previously established baseline when detecting phonemic, neologistic, and phonemic+neologistic paraphasias. 
We achieve AWERs of 45.0, 30.4, and 48.4 for phonemic, neologistic, and phonemic+neologistic paraphasias, respectively, which represents performance improvements over the previous approach of 16.9\%, 36.4\%, and 9.5\%.

When comparing the TD across seq2seq models, we observe a large performance gap between the use of MTL and STL objectives. 
These performance gaps demonstrate that STL models are not able to perform fine-grain paraphasia detection to the same degree as MTL models.
This performance gap is also likely a result of poor ASR performance, which can result in poor alignment and ultimately a larger TD metric.
These results suggest that in a seq2seq system, optimizing for both ASR performance and paraphasia detection simultaneously leads to better word-level paraphasia detection.

When looking at utterance-level average F1-scores, we find that the seq2seq paraphasia detection models outperform the previous state-of-the-art approach for phonemic, neologistic, and phonemic+neologistic paraphasias.
Our best-performing models achieve an F1 of 0.643, 0.688, and 0.706 for phonemic, neologistic, and phonemic+neologistic paraphasia detection, respectively. This represents performance improvements of 5.2\%, 13.9\%, and 18.9\% over the previous SOTA method.
When looking at the F1-scores for the seq2seq models we observe minimal performance changes when comparing different pretrained models and learning objectives.

\begin{figure}
  \centering
\includegraphics[width=\linewidth,height=0.3\textheight,keepaspectratio]{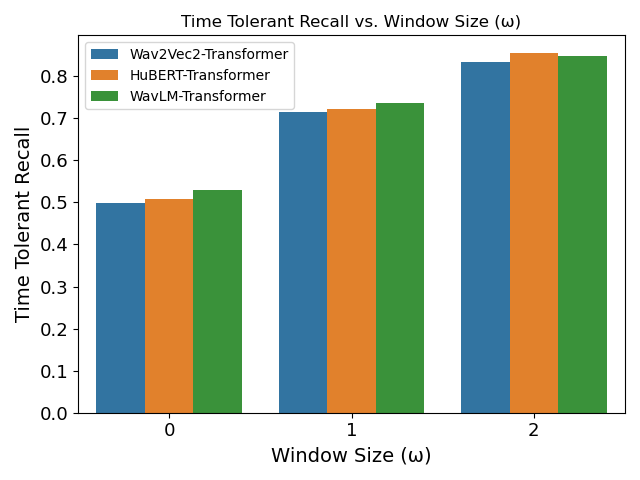}
  \caption{The impact of window size ($\omega$) on Time Tolerant Recall when using MTL seq2seq models.}
  \label{fig:word-level}
\end{figure}

Lastly, we explore the performance of word-level paraphasia detection using TTR and investigate the impact of window size ($\omega$) on seq2seq MTL models.
In Figure~\ref{fig:word-level}, we observe large TTR improvements as $\omega$ increases demonstrating that all these models exhibit high proximity to the ground truth label index. 
We note that WavLM-Transformer achieves the highest TTR when $\omega$=0, which could potentially occur due to better alignment as indicated by the higher ASR performance shown in Table~\ref{tab:asr}. 
However, as $\omega$ increases, we see that Hubert-Transformer closes the performance gap and outperforms WavLM-Transformer when $\omega$=2.
These results demonstrate that both HuBERT-Transformer and WavLM-Transformer approaches have high word-level paraphasia recall given a buffer size of a few words.

\noindent\textbf{Paraphasia Detection - Discussion} \\
We've shown that the proposed seq2seq model achieves better word-level and utterance-level paraphasia detection over prior work, highlighting the advantages of E2E training.
When comparing STL and MTL objectives we find that while the STL approach does demonstrate adequate utterance-level paraphasia detection, the performance of word-level paraphasia detection, as measured with AWER and TD, is poor. This is likely due to the final STL models having unoptimized ASR heads which in turn impair fine-grain word-level paraphasia detection.
In contrast, simultaneously optimizing both ASR and paraphasia detection tasks lead to models that are more robust interms of word-level and utterance-level paraphasia detection.
We find that the seq2seq models trained with MTL achieve the SOTA paraphasia detection for all word-level metrics and some utterance-level metrics.
With this in mind, we believe that the use of an MTL objective is essential for training paraphasia detection E2E.
Both the HuBERT-Transformer and WavLM-Transformer trained with MTL achieve high paraphasia detection performance depending on the paraphasia that is being detected. 
The HuBERT-Transformer performs better at phonemic paraphasia detection while the WavLM-Transformer performs better at neologistic paraphasia detection.
We find that either HuBERT-Transformer or WavLM-Transformer models are viable for detecting the presence of paraphasias in a given utterance and location of paraphasias within a few words.
When thinking about how automated paraphasia detection can be used to facilitate aphasia assessment, we believe that this work demonstrates the efficacy of a seq2seq model trained with MTL for both ASR and paraphasia detection.

\begin{table}[H]
\centering
\begin{tabular}{ |c || c | c | c | c |  }
 \toprule
  & 100 & 500 & 1000 & 2000 \\
 \midrule
    AWER & 49.9  & 49.9 & \textbf{37.7} & 51.8 \\
    TD & 8.8 & \textbf{8.1} & 8.5 & 8.7 \\
    F1-score & 0.68 & \textbf{0.70} & 0.68 & 0.69 \\
 \bottomrule
\end{tabular}
\vspace{5pt}
\caption{Analysis of tokenizer size on the MTL HuBERT-Transformer. All metrics were computed across all folds on the Scripts dataset.}
 \label{tab:vocab}
\end{table}


\subsection{Tokenizer Analysis}
Selecting a reasonable vocabulary size is critical in the subword tokenization process and can impact the performance of ASR and paraphasia detection. 
In this section, we explore the impact of vocabulary size on the performance of ASR and downstream paraphasia detection tasks. We use a SentencePiece tokenizer with a unigram tokenization scheme and sweep over vocabulary sizes of 100, 500, 1000, and 2000. We focus our analysis on the MTL WavLM-Transformer model for phonemic+neologistic paraphasias, which was one the best-performing model in section~\ref{sec:exp-PD}.

From table~\ref{tab:vocab}, we see that using a tokenizer size of 500 generally yields the best performance according to both word-level and utterance-level paraphasia detection metrics presented. The one metric that goes against this statement is AWER, where a tokenizer size of 1000 achieves a noticeably low AWER. We believe this could be due to the model converging at a point that is more optimal for ASR, ultimately resulting in a lower AWER.
We believe that for paraphasia detection selecting an appropriate tokenizer size is important as too large a vocabulary will result in sparse paraphasia labels, while too small a vocabulary can result in too much overlap between smaller sets of subtokens.

 \begin{table*}[h]
\centering
\begin{tabular}{ l | l  }
\toprule
    \multicolumn{2}{|c|}{P1\_B2\_SA\_C1-4} \\
    \midrule
    Ground Truth & fees/1 \hspace{2pt} speak/0 directing/0 to/0 me/0 and/0 din/1 me/0 time/0 to/0 myunikat/1 \\
    Predicted & please/1 meek/1 directly/0 to/0 me/0 and/0 then/1 me/1 time/1 to/0 myunikat/1 \\
    \midrule
    
    \multicolumn{2}{|c|}{P1\_T4\_SA\_C2-0} \\
    \midrule
    Ground Truth & I/0 han/1 asferaja/1 \\
    Predicted & I/0 have/1 afasa/1 \\
    \midrule

    \multicolumn{2}{|c|}{P3\_T4\_SA\_C3-1} \\
    \midrule
    Ground Truth & jersit/1 $<$eps$>$ means/0 I/0 have/0  diferkli/1 \hspace{5pt} vis/1 \hspace{5pt}  lanerj/1 \\
    Predicted & durs/1 \hspace{4pt} it/0  \hspace{5pt} means/0 I/0 have/0 diffritulti/1 landerj/1 $<$eps$>$\\
\bottomrule
\end{tabular}
\vspace{5pt}
\caption{Transcription Analysis: MTL Hubert-Transformer for phonemic and neologistic paraphasias}
\label{tab:transcription_verbatim}
\end{table*}

\subsection{Transcription Analysis}
Table~\ref{tab:transcription_verbatim} has some example AWER transcripts produced by the MTL HuBERT-Transformer model for phonemic and neologistic paraphasias. By examining some output, we can get a better understanding of ASR and paraphasia detection performance at the word-level.

For utterance P1\_B2\_SA\_C1-4, the alignment of the ground truth and predicted AWER transcripts is good. 
The model is able to correctly identify 3 paraphasias and produces 3 false positives. The strength of this model is the ability to detect the presence of paraphasias as well as which words they belong to. Additionally, the locations of the false positives are near the ground truth paraphasias, which can be acceptable for certain applications like flagging paraphasia regions. 
We also note that for word recognition, the ASR output does misrecognize some words.

For utterance P1\_T4\_SA\_C2-0, we can another case of good alignment and the paraphasia detection model is able to correctly predict all paraphasias.
We do see another instance though where the ASR head is unable to recognize the paraphasic words. This highlights the importance of the additional metrics like TD and TTR which focus the evaluation on just the paraphasia detection output.
This utterance highlights a common pattern of the model where the ASR error for paraphasic words is very high.
We believe this is in part due to the high variability of the pseudo-word targets for some paraphasias (described in section~\ref{sec:transcript-preprocess}).
This high label variability is also compounded by challenges such as high speaker variability and data scarcity, particularly for paraphasias.

For utterance P3\_T4\_SA\_C3-1, we can see some slight misalignment near the end of the transcript most notably for the word `lanerj'. 
The levenshtein distance that is used to align the AWER transcripts does not produce perfect alignments for evaluating word-level paraphasia detection.
This can produce misalignments highlighted in this example and ultimately motivates the use of metrics like TD and TTR that consider proximity.

\noindent\textbf{Transcription Analysis - Discussion} \\
This analysis highlights some of the strengths and weaknesses of the MTL HuBERT-Transformer model as well as some of the challenges associated with evaluating paraphasia detection.
We can see the model performing well when recognizing paraphasias at the word-level and generally good ASR performance.
However, one limitation is the poor ASR performance for paraphasic words, which we believe is due to the high variability with how these pseudo-word targets are generated.
Another challenge is the issue of misalignment that we see in utterance P3\_T4\_SA\_C3-1 for the paraphasia `lanerj'. 
This example highlights the importance of using metrics that take proximity into account, like TD and TTR, when evaluating word-level paraphasia detection.
These examples highlight the challenges associated with this task and the need for detailed evaluations that can help researchers better understand the strengths and limitations of the resulting machine-learning systems.

In clinical settings, these models can provide more feedback to medical professionals who are in the process of analyzing aphasic speech. 
One example of this is in streamlining the annotation process, where the model is used to flag paraphasic instances.
With applications like this in mind, slight misalignment issues can be overcome by the medical professional who has the context of both the recognized word and predicted paraphasia label (highlighted by the AWER transcript output).

\section{Conclusion}
This work investigates different methods for improving paraphasia detection, which can aid clinicians with traditional speech-language aphasic analyses and be helpful for specific treatment planning such as supplemental, self-driven, app-based therapy.
We first begin by evaluating existing ASR architectures for aphasic speech recognition.
We find that leveraging pretrained speech models is critical in low-resource domains such as aphasia and that fine-tuning with either an encoder-only or seq2seq architecture led to improved performance. 
Our best model is a seq2seq WavLM-Transformer model.

We then extend this approach and present a novel paraphasia detection model that is trained E2E and performs both speech recognition and binary paraphasia classification.
We explore the proposed seq2seq model with both MTL and STL objectives and compare against prior work on previously used word-level and utterance-level paraphasia detection metrics as well as provide additional follow-up evaluations for word-level paraphasia detection.
We demonstrate that either a HuBERT or WavLM seeded seq2seq model trained with MTL achieves state-of-the-art paraphasia detection performance at the word- and utterance-levels.
We provide some analyses on the effects of tokenizer size on paraphasia detection, which is a hyperparameter to consider for seq2seq models.
Lastly, we show some AWER output from MTL HuBERT-Transformer model to highlight some of the common strengths and weaknesses observed in the model and discuss how this could be used in clinical settings.

\section{Acknowledgements}
This research is based in part upon work supported by the National Science Foundation (NSF IIS-RI 2006618, NSF IIS-RI 008860, Graduate Research Fellowship Program). This research is supported in part through computational resources and services provided by Advanced Research Computing (ARC), a division of Information and Technology Services (ITS) at the University of Michigan, Ann Arbor.

\bibliographystyle{IEEEtran}
\bibliography{main}


\begin{IEEEbiography}
[{\includegraphics[width=1in,height=1.25in,clip,keepaspectratio]{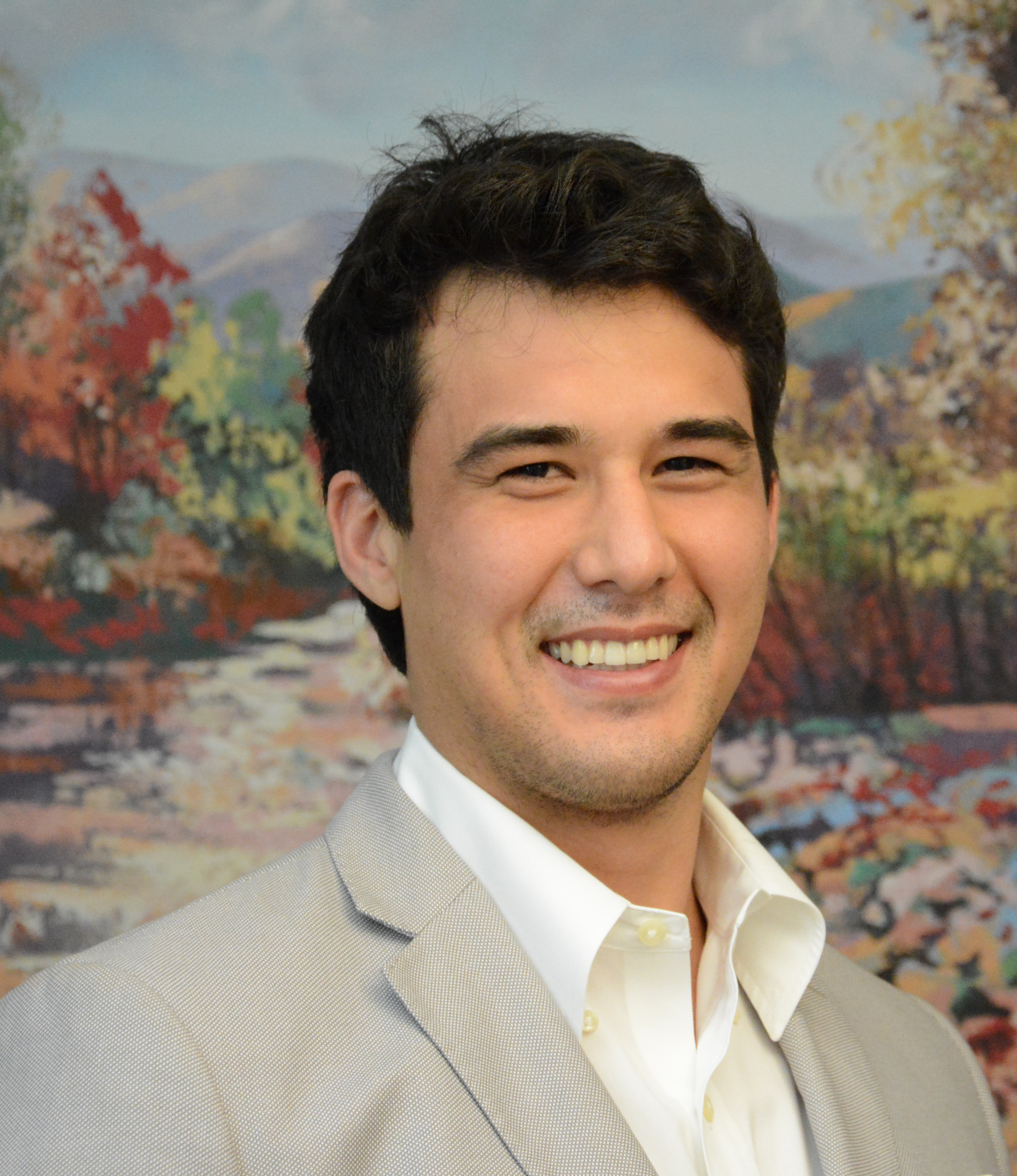}}]
{Matthew Perez} is a Ph.D. student at the University of Michigan working with Professor Emily Mower Provost. He received his B.S. degree in Computer Science from the University of Notre Dame in 2017 and his M.S. degree in Computer Science and Engineering from the University of Michigan in 2019. He was the recipient of the GEM Fellowship award (2019-2020) and the National Science Foundation Graduate Research Fellowship (2020-2023). His research interests include machine learning for speech-based assistive technology, automatic speech recognition, and computational paralinguistics. He is a member of IEEE and ISCA.
\end{IEEEbiography}

\begin{IEEEbiography}
[{\includegraphics[width=1in,height=1.25in,clip,keepaspectratio]{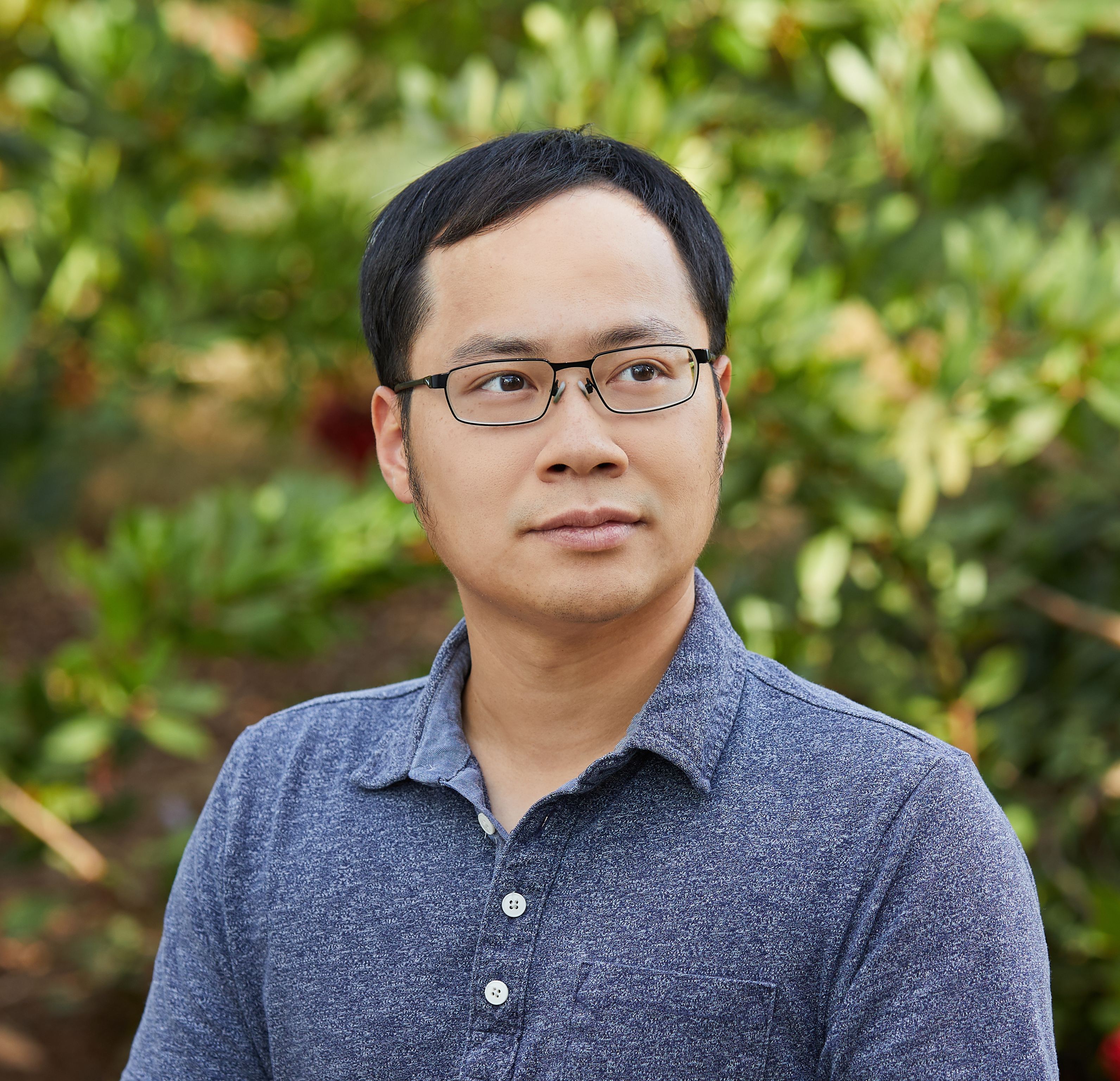}}]
{Duc Le} is a Research Lead at ByteDance, where he works on end-to-end music generation and large-scale music understanding. Previously, he was a Research Scientist Manager at Meta, specializing in automatic speech recognition and spoken language understanding. He received his B.S. in Computer Science (summa cum laude) from the University of Texas at Dallas, Richardson, TX in 2012 and his M.S. and Ph.D. in Computer Science from the University of Michigan, Ann Arbor, MI in 2014 and 2017, respectively. He is a member of IEEE and ISCA, and has served as session chair for ICASSP and INTERSPEECH.
\end{IEEEbiography}

\begin{IEEEbiography}
[{\includegraphics[width=1in,height=1.25in,clip,keepaspectratio]{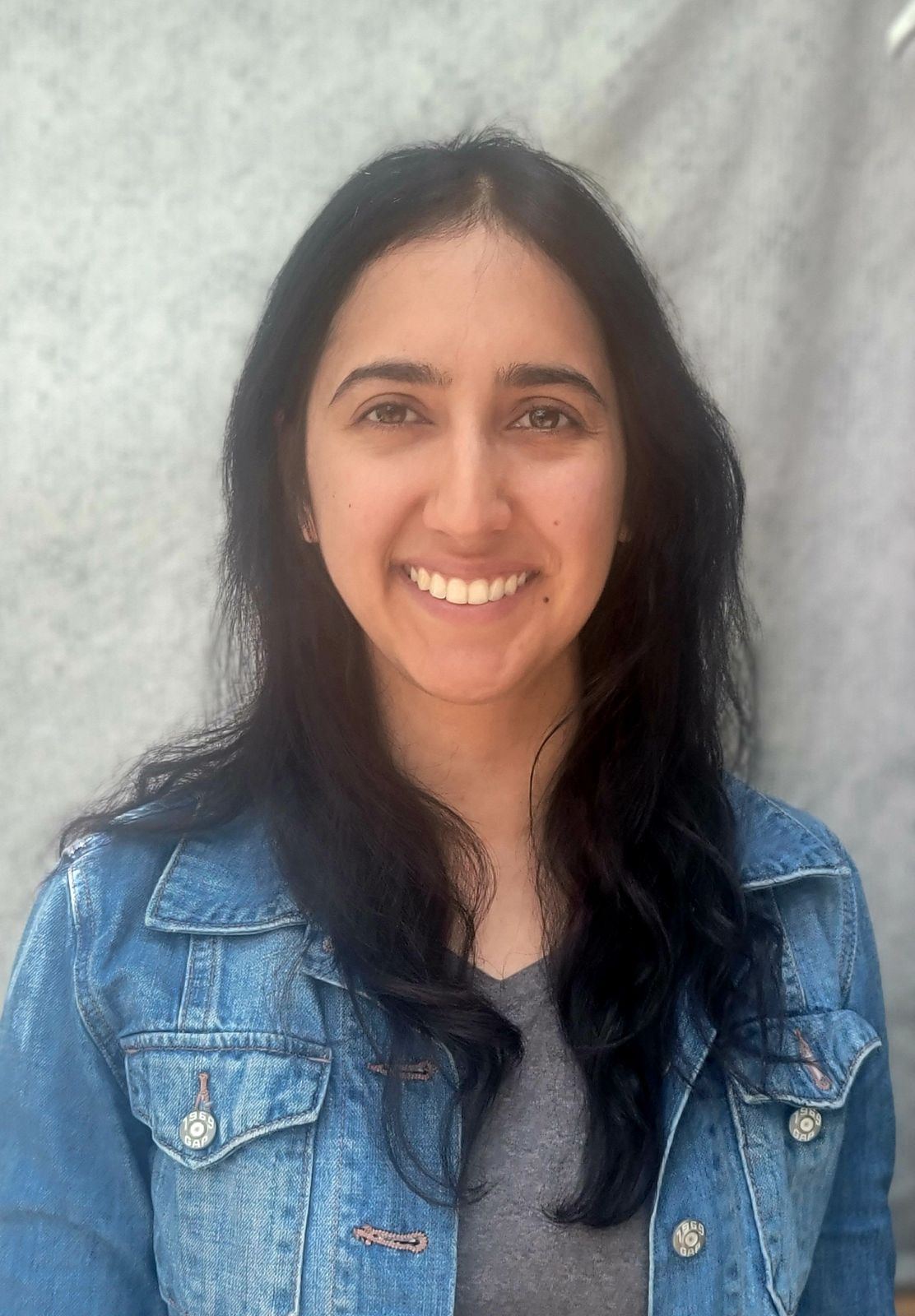}}]
{Amrit Romana} received her B.S. degree in Mathematics, and M.S. degree in Computer Science and Engineering from University of Michigan, in 2014 and 2020, respectively. She is currently a Ph.D. student in Computer Science and Engineering at University of Michigan. Her research interests include speech processing and machine learning with the goal of making speech-based technologies more accessible. She is a member of IEEE and ISCA.
\end{IEEEbiography}

\begin{IEEEbiography}
[{\includegraphics[width=1in,height=1.25in,clip,keepaspectratio]{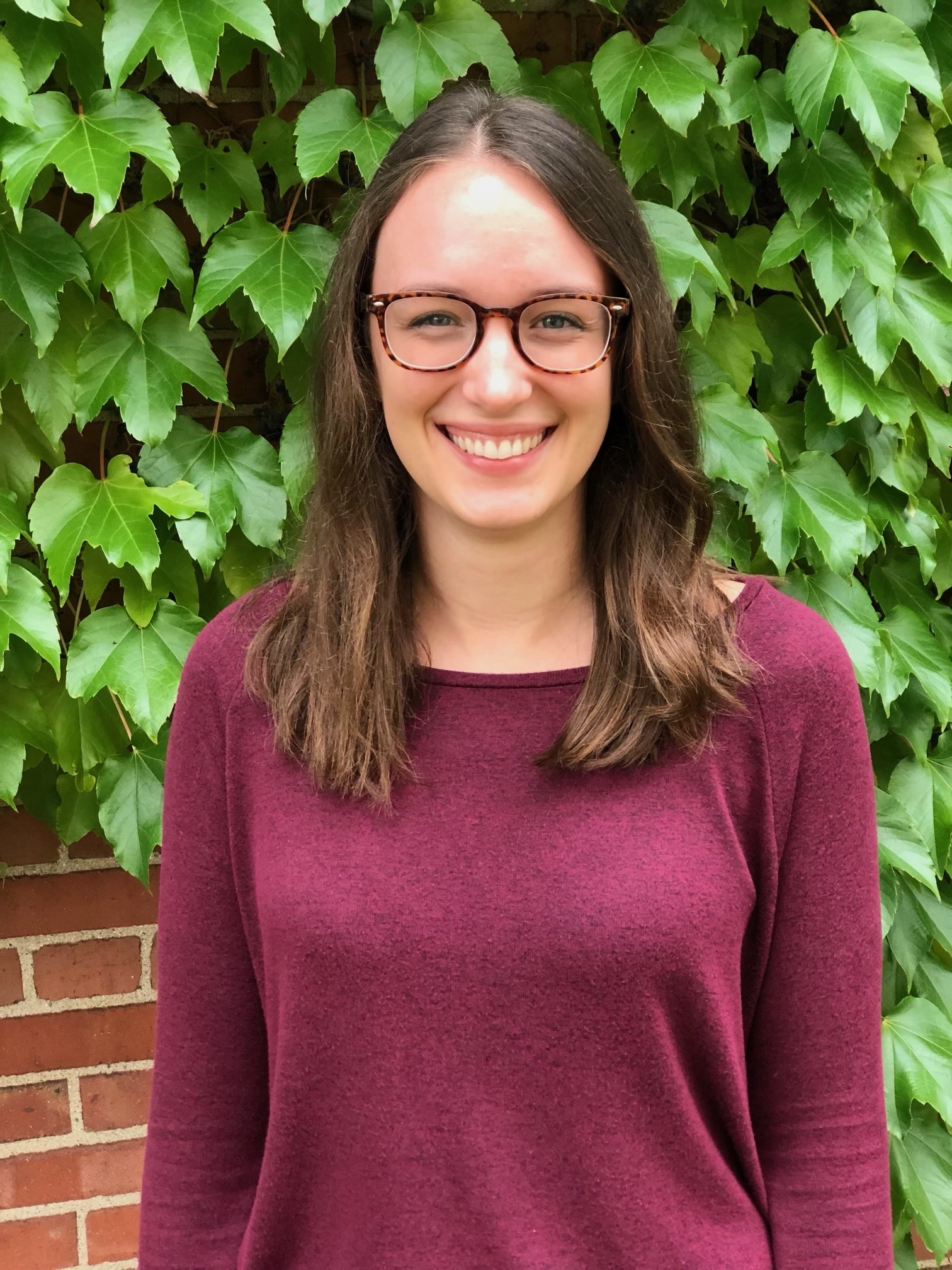}}]
{Elise Jones} is a Senior Speech-Language Pathologist at the University Center for Language and Literacy at the University of Michigan. She earned her B.S. in Speech and Hearing Sciences from Purdue University in 2014 and her M.A. in Speech-Language Pathology from Indiana University in 2016. She holds a Certificate of Clinical Competence through the American Speech and Hearing Association (ASHA), primarily providing assessment and treatment for adults with aphasia through the University of Michigan Aphasia Program. Her clinical interests include supporting individuals with aphasia and their families in communication, engagement, and improving overall independence and quality of life. 
\end{IEEEbiography}

\begin{IEEEbiography}
[{\includegraphics[width=1in,height=1.25in,clip,keepaspectratio]{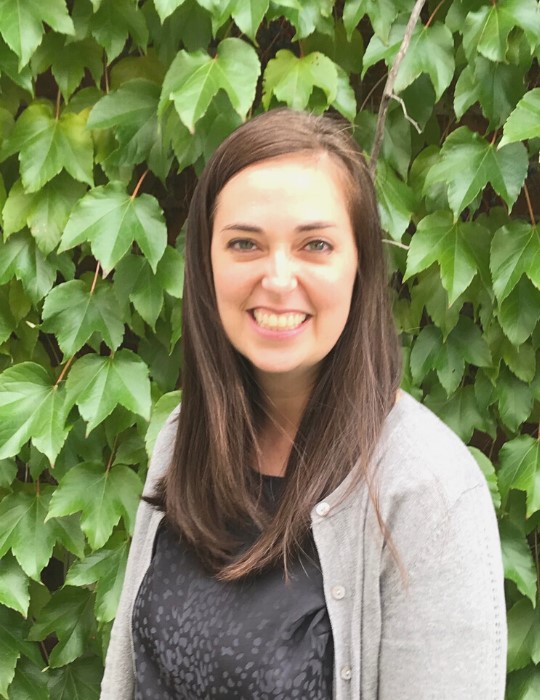}}]
{Keli Licata} is a Senior Speech-Language Pathologist and the Education Coordinator at the University Center for Language and Literacy (UCLL) at the University of Michigan. She works primarily with individuals with aphasia and their caregivers within the University of Michigan Aphasia Program (UMAP). She earned her B.A. in Linguistics and Psychology from the University of Michigan and her M.A. in Speech-Language Pathology from Indiana University. Keli holds a Certificate of Clinical Competence from the American Speech-Language-Hearing Association (ASHA) and is licensed to provide teletherapy services across multiple states. Her clinical areas of interest include providing evidence-based, person- and family-centered care for adults with aphasia. Her collaborative research interests currently include characterizing the stress experienced by caregivers for adults with aphasia. Keli will also begin teaching the graduate-level aphasia course in the Department of Communication Sciences and Disorders at Michigan State University in January, 2024.
\end{IEEEbiography}

\begin{IEEEbiography}
[{\includegraphics[width=1in,height=1.25in,clip,keepaspectratio]{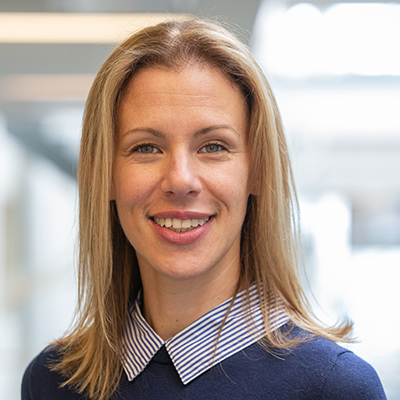}}]
{Emily Mower Provost} (M’11, SM'17) is a Professor in Computer Science and Engineering at the University of Michigan. She received her Ph.D. in Electrical Engineering from the University of Southern California (USC), Los Angeles, CA in 2010. She is a Toyota Faculty Scholar (2020) and has been awarded a National Science Foundation CAREER Award (2017), the Oscar Stern Award for Depression Research (2015), a National Science Foundation Graduate Research Fellowship (2004-2007).  She is an Associate Editor for IEEE Transactions on Affective Computing and the IEEE Open Journal of Signal Processing.  She has also served as Associate Editor for Computer Speech and Language and ACM Transactions on Multimedia.  She has received best paper awards or finalist nominations for Interspeech 2008, ACM Multimedia 2014, ICMI 2016, and IEEE Transactions on Affective Computing.  Among other organizational duties, she has been Program Chair for ACII (2017, 2021), ICMI (2016, 2018).  Her research interests are in human-centered speech and video processing, multimodal interfaces design, and speech-based assistive technology. The goals of her research are motivated by the complexities of the perception and expression of human behavior. 
\end{IEEEbiography}

\vfill

\end{document}